\newcommand{\CLASSINPUTtoptextmargin}{19mm}%
\newcommand{\CLASSINPUTbottomtextmargin}{43mm}%
\newcommand{\CLASSINPUTinnersidemargin}{12.9mm}%
\newcommand{\CLASSINPUToutersidemargin}{12.9mm}%
\documentclass[conference,10pt,a4paper]{IEEEtran}
\usepackage{amsmath}
\usepackage{xcolor}
\usepackage{times}
\usepackage{graphicx}
\usepackage{multirow}
\usepackage{siunitx}[mode=match]
\usepackage{hhline}
\usepackage{amsmath}
\usepackage[none]{hyphenat}
\usepackage{float}
\usepackage{subfig}
\usepackage{iftex}
\usepackage{amssymb,mathtools}
\usepackage{algorithm}
\usepackage{algpseudocode}   
\usepackage{pgfplots} 
\usepackage{circuitikz}[european] 
\usepackage[percent]{overpic}
\usepackage{graphicx}
\usepackage{amsmath}
\usepackage{psfrag}
\usepackage{subfig}
\usepackage{acronym}
\usepackage[font=footnotesize]{caption}
\usepackage{xcolor}     
\usepackage{subcaption} 
\usepackage{trfsigns} 
\usepackage{nohyperref}  

\newacro{3GPP}{3rd Generation Partnership Project}
\newacro{5G}{fifth generation}
\newacro{5G NR}{Fifth Generation New Radio}
\newacro{6G}{sixth generation}
\newacro{A/D}{analog-to-digital}
\newacro{AAL}{array aperture line}
\newacro{ABE}{analog back-end}
\newacro{ADC}{analog-to-digital converter}
\newacro{ADS}{Advanced Design System} 
\newacro{AFE}{analog front-end}
\newacro{AGV}{automatic guided vehicle}
\newacro{AM-AM}{amplitude-to-amplitude modulation}
\newacro{AM-PM}{amplitude-to-phase modulation}
\newacro{AWGN}{additive white Gaussian noise}
\newacro{B5G}{beyond \ac{5G}}
\newacro{BB}{baseband}
\newacro{BER}{bit error ratio}
\newacro{BPSK}{binary phase-shift keying}
\newacro{BP}{band-pass}
\newacro{BS}{base station}
\newacro{CDM}{code-division multiplexing}
\newacro{CFO}{carrier frequency offset}
\newacro{CFR}{channel frequency response}
\newacro{CIR}{channel impulse response}
\newacro{CoMP}{coordinated multipoint}
\newacro{CP}{cyclic prefix}
\newacro{CPE}{common phase error}
\newacro{CPO}{carrier phase offset}
\newacro{CRLB}{Cram\'er-Rao lower bound}
\newacro{CS}{chirp sequence}
\newacro{CSI}{channel state information}
\newacro{CW}{continuous wave}
\newacro{CZT}{chirp Z-transform}
\newacro{D/A}{digital-to-analog}
\newacro{DAC}{digital-to-analog converter}
\newacro{DC}{direct current}
\newacro{DDC}{digital down-conversion}
\newacro{DDS}{direct digital synthesis}
\newacro{DFRC}{dual-function radar-communication or dual-functional radar-communication}
\newacro{DFnT}{discrete Fresnel transform}
\newacro{DFT}{discrete Fourier transform}
\newacro{DL}{downlink}
\newacro{DMRS}{demodulation reference signal}
\newacro{DoA}{direction-of-arrival}
\newacro{DoD}{direction-of-departure}
\newacro{DPD}{digital pre-distortion}
\newacro{DUC}{digital up-conversion}
\newacro{ETSI}{European Telecommunications Standards Institute}
\newacro{EVM}{error vector magnitude}
\newacro{FDE}{frequency-domain equalization}
\newacro{FDM}{frequency-division multiplexing}
\newacro{FO}{frequency offset}
\newacro{FR2}{Frequency Range 2}
\newacro{gNB}{gNodeB}
\newacro{HP}{high-pass}
\newacro{IBFD}{in-band full duplex}
\newacro{ICI}{intercarrier interference}
\newacro{IDFT}{inverse discrete Fourier transform}
\newacro{IDFnT}{inverse discrete Fresnel transform}
\newacro{IF}{intermediate frequency}
\newacro{IHE}{Institute of Radio Frequency Engineering and Electronics}
\newacro{I/Q}{in-phase/quadrature}
\newacro{IBO}{input back-off}
\newacro{IP1dB}{input-referred 1-dB compression point}
\newacro{ISAC}{integrated sensing and communication}
\newacro{ISI}{intersymbol interference}
\newacro{ISLR}{integrated-sidelobe level ratio}
\newacro{IM3}{third-order intermodulation}
\newacro{IoT}{Internet of Things}
\newacro{JCAS}{joint communication and sensing}
\newacro{KIT}{Karlsruhe Institute of Technology}
\newacro{KPI}{key performance indicator}
\newacro{LDPC}{low-density parity-check}
\newacro{LFSR}{linear-feedback shift register}
\newacro{LNA}{low-noise amplifier}
\newacro{LTE}{long term evolution}
\newacro{LO}{local oscillator}
\newacro{LoS}{line-of-sight}
\newacro{LP}{low-pass}
\newacro{LPI}{low probability of intercept}
\newacro{LS}{least squares}
\newacro{mmWave}{milimeter wave}
\newacro{MIMO}{multiple-input multiple-output}
\newacro{MLE}{maximum likelihood estimator}
\newacro{MLS}{maximum-length sequence}
\newacro{MRC}{maximal-ratio combining}
\newacro{MUSIC}{multiple signal classification}
\newacro{IMD}{intermodulation distortion}
\newacro{NAF}{normalized angular frequency}
\newacro{NB}{narrowband}
\newacro{NLoS}{non-line-of-sight}
\newacro{NR}{new radio}
\newacro{OCDM}{orthogonal chirp-division multiplexing}
\newacro{OFDM}{orthogonal frequency-division multiplexing}
\newacro{OOB}{out-of-band}
\newacro{OTA}{over-the-air}
\newacro{P/S}{parallel-to-serial}
\newacro{PA}{power amplifier}
\newacro{PACF}{periodic autocorrelation function}
\newacro{PAPR}{peak-to-average power ratio}
\newacro{PCCF}{periodic cross-correlation function}
\newacro{PLC}{powerline communication}
\newacro{PLL}{phase-locked loop}
\newacro{PMCW}{phase-modulated continuous wave}
\newacro{PMN}{perceptive mobile network}
\newacro{PN}{oscillator phase noise}
\newacro{PoC}{proof-of-concept}
\newacro{PPLR}{peak power loss ratio}
\newacro{PRBS}{pseudorandom binary sequence}
\newacro{PRS}{positioning reference signal}
\newacro{PSD}{power spectral density}
\newacro{PSF}{point spread function}
\newacro{PSLR}{peak-to-sidelobe level ratio}
\newacro{QPSK}{quadrature phase-shift keying}
\newacro{RadCom}{radar-communication}
\newacro{RCS}{radar cross section}
\newacro{RF}{radio-frequency}
\newacro{RFS}{random finite set}
\newacro{RIS}{reflective intelligent surface}
\newacro{RMS}{root mean square}
\newacro{RMSE}{root mean squared error}
\newacro{RX}{receiver}
\newacro{SC}[S\&C]{Schmidl \& Cox}
\newacro{SFO}{sampling frequency offset}
\newacro{SIC}{self-interference cancellation}
\newacro{SINR}{signal-to-interference-plus-noise ratio}
\newacro{SIR}{signal-to-interference ratio}
\newacro{SISO}{single-input single-output}
\newacro{SJ}{sampling jitter}
\newacro{SNR}{signal-to-noise ratio}
\newacro{SoC}{system-on-a-chip}
\newacro{SQNR}{signal-to-quantization-noise ratio}
\newacro{SSB}{synchronization signal block}
\newacro{STO}{symbol time offset}
\newacro{S/P}{serial-to-parallel}
\newacro{TDD}{time-division duplexing}
\newacro{TDE}{time-domain equalization}
\newacro{TDM}{time-division multiplexing}
\newacro{TDR}{time-domain reflectometry}
\newacro{TITO}{tilt inference of time offset}
\newacro{TO}{time offset}
\newacro{TR}{technical report}
\newacro{TS}{technical specification}
\newacro{TX}{transmitter}
\newacro{UE}{user equipment}
\newacro{UL}{uplink}
\newacro{ULA}{uniform linear array}
\newacro{V2V}{vehicle-to-vehicle}
\newacro{ZF}{zero forcing}
\newacro{ZP}{zero padding}

\ifPDFTeX%
\usepackage{t1enc}
\usepackage{times}

\fi%
\ifXeTeX%
\usepackage{fontspec}
\fi%
\ifLuaTeX%
\usepackage{fontspec}
\fi%
\makeatletter

\def\@maketitle{\newpage
\bgroup\par\addvspace{0.5\baselineskip}\centering%
\ifCLASSOPTIONtechnote
   {\bfseries\large\@IEEEcompsoconly{\sffamily}\@title\par}\vskip 1.3em{\lineskip .5em\@IEEEcompsoconly{\sffamily}\@author
   \@IEEEspecialpapernotice\par{\@IEEEcompsoconly{\vskip 1.5em\relax
   \@IEEEtitleabstractindextextbox{\@IEEEtitleabstractindextext}\par
   \hfill\@IEEEcompsocdiamondline\hfill\hbox{}\par}}}\relax
\else
   \vskip0.2em{\EuMWtitlesize\ifCLASSOPTIONtransmag\bfseries\LARGE\fi\@IEEEcompsoconly{\sffamily}\@IEEEcompsocconfonly{\normalfont\normalsize\vskip 2\@IEEEnormalsizeunitybaselineskip
   \bfseries\Large}\@title\par}\vskip1.0em\par
   \ifCLASSOPTIONconference%
      {\@IEEEspecialpapernotice\mbox{}\vskip\@IEEEauthorblockconfadjspace%
       \mbox{}\hfill\begin{@IEEEauthorhalign}\@author\end{@IEEEauthorhalign}\hfill\mbox{}\par}\relax
   \else
      \ifCLASSOPTIONpeerreviewca
         {\@IEEEcompsoconly{\sffamily}\@IEEEspecialpapernotice\mbox{}\vskip\@IEEEauthorblockconfadjspace%
          \mbox{}\hfill\begin{@IEEEauthorhalign}\@author\end{@IEEEauthorhalign}\hfill\mbox{}\par
          {\@IEEEcompsoconly{\vskip 1.5em\relax
           \@IEEEtitleabstractindextextbox{\@IEEEtitleabstractindextext}\par\hfill
           \@IEEEcompsocdiamondline\hfill\hbox{}\par}}}\relax
      \else
         \ifCLASSOPTIONtransmag
           {\@IEEEspecialpapernotice\mbox{}\vskip\@IEEEauthorblockconfadjspace%
            \mbox{}\hfill\begin{@IEEEauthorhalign}\@author\end{@IEEEauthorhalign}\hfill\mbox{}\par
           {\vspace{0.5\baselineskip}\relax\@IEEEtitleabstractindextextbox{\@IEEEtitleabstractindextext}\vspace{-1\baselineskip}\par}}\relax
         \else
           {\lineskip.5em\@IEEEcompsoconly{\sffamily}\sublargesize\@author\@IEEEspecialpapernotice\par
           {\@IEEEcompsoconly{\vskip 1.5em\relax
            \@IEEEtitleabstractindextextbox{\@IEEEtitleabstractindextext}\par\hfill
            \@IEEEcompsocdiamondline\hfill\hbox{}\par}}}\relax
         \fi
      \fi
   \fi
\fi\par\addvspace{0.0\baselineskip}\egroup}

\def\EuMWtitlesize{\@setfontsize{\EuMWtitlesize}{24}{24pt}}
\def\EuMWauthorsize{\@setfontsize{\EuMWauthorsize}{11}{11pt}}
\def\EuMWaffilsize{\@setfontsize{\EuMWaffilsize}{10}{10pt}}
\def\EuMWcaptionsize{\@setfontsize{\EuMWcaptionsize}{9}{10pt}}
\def\EuMWbibsize{\@setfontsize{\EuMWbibsize}{8}{10pt}}

\def\@IEEEauthorblockNstyle{\EuMWauthorsize\@IEEEcompsocnotconfonly{\sffamily}\@IEEEcompsocconfonly{\large}}
\def\@IEEEauthorblockAstyle{\EuMWaffilsize\@IEEEcompsocnotconfonly{\sffamily}\@IEEEcompsocconfonly{\itshape}\@IEEEcompsocconfonly{\large}}
\def\@IEEEauthordefaulttextstyle{\EuMWauthorsize\@IEEEcompsocnotconfonly{\sffamily}\sublargesize}

\def\thebibliography#1{\section*{\refname}%
    \addcontentsline{toc}{section}{\refname}%
    \EuMWbibsize\@IEEEcompsocconfonly{\small}\vskip 0.3\baselineskip plus 0.1\baselineskip minus 0.1\baselineskip
    \list{\@biblabel{\@arabic\c@enumiv}}%
    {\settowidth\labelwidth{\@biblabel{#1}}%
    \leftmargin\labelwidth
    \advance\leftmargin\labelsep\relax
    \itemsep \IEEEbibitemsep\relax
    \usecounter{enumiv}%
    \let\p@enumiv\@empty
    \renewcommand\theenumiv{\@arabic\c@enumiv}}%
    \let\@IEEElatexbibitem\bibitem%
    \def\bibitem{\@IEEEbibitemprefix\@IEEElatexbibitem}%
\def\newblock{\hskip .11em plus .33em minus .07em}%
\ifCLASSOPTIONtechnote\sloppy\clubpenalty4000\widowpenalty4000\interlinepenalty100%
\else\sloppy\clubpenalty4000\widowpenalty4000\interlinepenalty500\fi%
    \sfcode`\.=1000\relax}

\long\def\@makecaption#1#2{%
\ifx\@captype\@IEEEtablestring%
\par\@IEEEtabletopskipstrut
\else
\@IEEEfigurecaptionsepspace
\fi
\setbox\@tempboxa\hbox{\normalfont\footnotesize {#1.}\nobreakspace\nobreakspace #2}%
\ifdim \wd\@tempboxa >\hsize%
\setbox\@tempboxa\hbox{\normalfont\footnotesize {#1.}\nobreakspace\nobreakspace}%
\parbox[t]{\hsize}{\normalfont\footnotesize\noindent\unhbox\@tempboxa#2}%
\else
\ifCLASSOPTIONconference \hbox to\hsize{\normalfont\footnotesize\hfil\box\@tempboxa\hfil}%
\else \hbox to\hsize{\normalfont\footnotesize\box\@tempboxa\hfil}%
\fi\fi
\ifx\@captype\@IEEEtablestring%
\@IEEEtablecaptionsepspace
\else
\fi}

\newlength\tablecaptiontotableskip
\newlength\figuretocaptionskip
\def\@IEEEfigurecaptionsepspace{\vskip\figuretocaptionskip\relax}%
\def\@IEEEtablecaptionsepspace{\vskip\tablecaptiontotableskip\relax}%

\def\abstract{\normalfont%
\@IEEEabskeysecsize\bfseries\textit{\abstractname}\,\bfseries\textit{---}\,%
\@IEEEgobbleleadPARNLSP}%

\def\IEEEkeywords{\normalfont%
\@IEEEabskeysecsize\bfseries\textit{\IEEEkeywordsname}\,\bfseries\textit{---}\,%
\@IEEEgobbleleadPARNLSP}%
\def\endIEEEkeywords{\relax\vspace{0.67ex}%
\par\if@twocolumn\else\endquotation\fi%
\normalsize\normalfont}%

\def\@IEEEauthorblockNtopspace{0ex}
\def\@IEEEauthorblockAtopspace{1mm}
\def\tablename{Table}
\def\thetable{\arabic{table}}
\def\IEEEkeywordsname{Keywords}
\def\subsubsection{\@startsection{subsubsection}{3}{\z@}{1.5ex plus 1.5ex minus 0.5ex}%
{0.7ex plus .5ex minus 0ex}{\normalfont\normalsize\itshape}}%
\newlength{\CPheadmatchindent}%
\def\@seccntformat#1{\hbox to\CPheadmatchindent{\csname the#1dis\endcsname}\hskip 0.1em \relax}
\IEEEilabelindentA \parindent
\IEEEilabelindent \IEEEilabelindentA
\IEEEelabelindent \parindent
\IEEEdlabelindent \parindent
\IEEElabelindent \parindent
\makeatother

\usepackage{fancyhdr}
\fancypagestyle{empty}{fancy}

\begin{document}
\raggedbottom
%
%
%
\title{Optimization-Based Behavioral Modeling of Mixers for Frequency Comb OFDM Radar Processing}
%
%
\author{%
\IEEEauthorblockN{%
Umut Utku Erdem\IEEEauthorrefmark{1}, 
Henning Poensgen\IEEEauthorrefmark{1}, 
Taewon Jeong\IEEEauthorrefmark{1},
Lucas Giroto\IEEEauthorrefmark{1}\IEEEauthorrefmark{2},
Benjamin Nuss\IEEEauthorrefmark{3},\\
{\.I}brahim Ka\u{g}an Aksoyak\IEEEauthorrefmark{4},
Ahmet Cagri Ulusoy\IEEEauthorrefmark{1},
Thomas Zwick\IEEEauthorrefmark{1}
}
\IEEEauthorblockA{%
\IEEEauthorrefmark{1}Institute of Radio Frequency Engineering and Electronics, Karlsruhe Institute of Technology, Germany\\
\IEEEauthorrefmark{2}Nokia Bell Labs Stuttgart, Germany\\
\IEEEauthorrefmark{3}Professorship of Microwave Sensors and Sensor Systems, Technical University of Munich, Germany\\
\IEEEauthorrefmark{4}Kilby Labs, Texas Instruments, Dallas, TX, USA\\
E-mail: umut.erdem@kit.edu}
}
%
\maketitle
%
%
%
\begin{abstract}
This paper presents an optimization-based behavioral model for mixers driven by multi-tone \ac{LO} signals, considered specifically for frequency comb \ac{OFDM} radar applications. Unlike traditional models, the proposed approach is designed and tested for multi-tone \ac{LO} excitations. The model uses polynomial nonlinearities for both \ac{IF} and \ac{LO} ports, supported by spectrum-domain fitting that selectively emphasizes strong intermodulation products. In addition, a polynomial block is introduced to capture input power-dependent phase nonlinearity. The approach is validated using circuit-level simulations and supported by measurements. Radar processing results show the model replicates distortive effects in simulations. The proposed model enables rapid system-level performance estimations and waveform optimization, replacing computationally expensive circuit-level simulations.
\end{abstract}
%
%

\section{Introduction}

The accurate behavioral modeling of mixers plays a vital role in determining the overall performance of modern \ac{RF} systems. In high-frequency applications, especially those involving high \ac{PAPR} and wideband signals, mixers lead to strong \ac{AM-AM} and \ac{AM-PM} distortions, harmonic generation, and \ac{IMD}, which necessitates accurate behavioral modeling.
Traditional mixer models have primarily focused on \ac{AM-AM} distortion using polynomial blocks around ideal mixer \cite{Vandermot}. More recent approaches separate amplitude, phase and memory effects, enabling better characterization of mixer behavior \cite{Multibox}. These models are crucial during the performance investigations of systems operating near 1-dB compression point ($\mathrm{P}_{\qty{1}{dB}}$).

The increasing interest in \ac{ISAC} for future \ac{6G} cellular standards and beyond \cite{survey_ISAC} has led to a paradigm shift in waveform design and system architecture.
In this context, \ac{OFDM} is a suitable multicarrier modulation scheme due to its high spectral efficiency and compatibility with existing communication standards, besides robust sensing performance.

To enable high-resolution \ac{OFDM}-based radar sensing, however, high \ac{ADC} and \ac{DAC} sampling rates are required. A possible solution is frequency comb \ac{OFDM} radar \cite{Nuss_comb}, which uses multiple \ac{LO} tones to upconvert a sparse low-rate \ac{OFDM} baseband signal, expanding effective bandwidth while relaxing the baseband sampling requirements. Unlike conventional \ac{OFDM}, where a single \ac{LO} tone is used, frequency comb \ac{OFDM} poses new challenges at the mixer stage. Existing frequency comb \ac{OFDM} systems \cite{Nuss_comb} use multiple parallel mixers driven by isolated \ac{LO} tones, increasing effective bandwidth but at the cost of increased hardware. This parallelization makes the mixer stage trackable. However, a more practical approach would be using multiple \ac{LO} tones with a single mixer. Utilizing several \ac{LO} tones within a single mixer introduces additional challenges, which can affect the system performance unless properly investigated.

In this work, we propose a novel optimization-based behavioral model customized for mixers under multi-tone \ac{LO} excitation, directly aligned with frequency comb OFDM radar applications. The proposed model extends previous behavioral modeling frameworks in \cite{Multibox} by considering spectrum-domain fitting and customizes multibox model for double-balanced mixers. The approach is validated through \ac{ADS} simulations, which are verified through measurements.
By replacing \ac{ADS} simulations with the proposed efficient behavioral block, the framework enables rapid waveform design and system validation.
\section{Custom Gilbert-Cell Mixer Design}
A double-balanced Gilbert-cell mixer was designed and fabricated to support frequency comb OFDM radar operation, specifically targeting scenarios with multi-tone \ac{LO} excitation and wideband signals. The behavioral modeling developed in this work is primarily based on simulations. The circuit itself is implemented in IHP's SG13G2Cu technology.

The complete schematic, including bias circuitry, is shown in Fig.~\ref{fig:mixer_schem}. Because the Gilbert-cell inherently suppresses even-order distortion products due to its differential nature, it simplifies the behavioral model optimization process. The mixer core is designed symmetrically, following the approach of \cite{Aksoyak2024}. 
All differential ports are \ac{DC} coupled to support true baseband signal excitation.

The $g_m$-stage transistors $Q_{1,2}$ consist of 10 fingers, $\qty{0.07}{\micro\meter} \times \qty{0.09}{\micro\meter}$ each. The switching quad $Q_{3-6}$ are half that size with 5 fingers. The biasing is ensured by a current mirror, employing $Q_{7,8}$ with each utilizing just one finger. A current density of \qty{2}{\milli\ampere} per finger results in a total power consumption of \qty{160}{\milli\watt}. The operating point is optimized for maximum conversion gain and an output compression point of $\mathrm{OP}_{\qty{1}{dB}} \geq \qty{0}{dBm}$. To enhance input-referred linearity, bandwidth, and matching, resistive emitter degeneration ($R_E$) parallel termination resistors ($R_{1-4}$) are employed at all ports, at the cost of increased power consumption. The manufactured chip is shown in Fig.~\ref{fig:mixer_micrograph}. 
The mixer has been characterized and supports single-tone \ac{LO} excitation from $9$ to $\SI{14}{\giga\hertz}$ with a small-signal \ac{IF}  bandwidth from DC to $\qty{7.5}{\giga\hertz}$ at a nominal single-tone \ac{LO} power of $P_\mathrm{LO} = \qty{0}{dBm}$. Further information about compression and linearity can be found in Sec.~\ref{meas_results}.

\pgfdeclarelayer{background}
\pgfdeclarelayer{middle}
\pgfdeclarelayer{foreground}
\pgfsetlayers{background,middle,main,foreground}

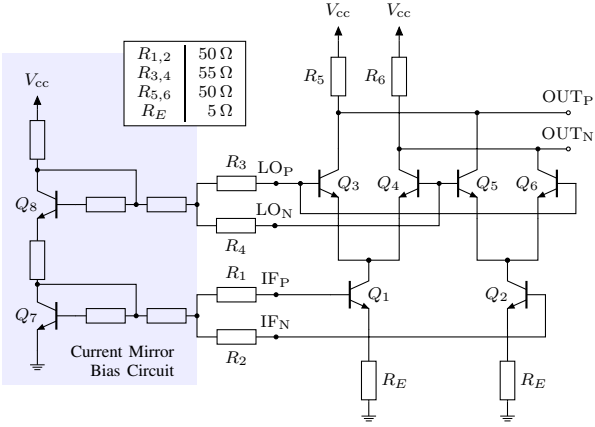
\begin{figure}[t]
    \centering
    \begin{circuitikz}[scale=0.8,transform shape,line width=0.5pt, european]
	\ctikzset{
		font = \footnotesize,
		bipoles/thickness=1,
		bipoles/length=0.8cm,
		capacitors/scale=0.75,
		tripoles/mos style/arrows,
		transistors/arrow pos=end,
		transistors/scale=1.3,
		grounds/thickness=0.85,
		flipflops/thickness=1,
		flipflops/scale=1,
		annotation distance = -3pt,
		monopoles/vcc/arrow={Triangle[width=0.8*\scaledwidth, length=\scaledwidth]},
		amplifiers/thickness=1,
		amplifiers/scale=2,
		amplifiers/plus={},
		amplifiers/minus={}
	}
        \begin{pgfonlayer}{middle}
            \draw 
            (0,0) node[above]{$\mathrm{IF_P}$} coordinate(IFPinput)
                to [short,*-] ++(0.9,0)
                node[npn,anchor=B] (Q1) {}
            (Q1.center) node[anchor=west,xshift=-4pt]{$Q_1$}
            (Q1.E)
                to [open] ++(1.14,0) coordinate(if_gnd_diff)
            (Q1.E) 
                to [short] ++(0,-0.5) 
                to [R=$R_{E}$] ++(0,-0.7)
                node[ground]{}
            (if_gnd_diff) 
                to [open] ++(1.14,0)
                node[npn,anchor=E,xscale=-1] (Q2){}
                (Q2.center) node[anchor=east,xshift=4pt]{$Q_2$}
            (Q2.B) |- (0.5,-0.7)
                to[short,-*] ++(-0.5,0) coordinate(IFNinput)
                node[above]{$\mathrm{IF_N}$}
            (Q1.C) 
                to [short,*-] ++(-0.5,0)
                to [short,-] ++(0,0.7)
                node[npn,anchor=E] (Q3){}
                (Q3.center) node[anchor=west,xshift=-4pt]{$Q_3$}
            (Q1.C) 
                to [short,*-] ++(0.5,0)
                to [short,-] ++(0,0.7)
                node[npn,anchor=E,xscale=-1] (Q4){}
                (Q4.center) node[anchor=east,xshift=4pt]{$Q_4$}
            (Q2.C) 
                to [short,*-] ++(-0.5,0)
                to [short,-] ++(0,0.7)
                node[npn,anchor=E] (Q5){}
                (Q5.center) node[anchor=west,xshift=-4pt]{$Q_5$}
            (Q2.C) 
                to [short,*-] ++(0.5,0)
                to [short,-] ++(0,0.7)
                node[npn,anchor=E,xscale=-1] (Q6){}
                (Q6.center) node[anchor=east,xshift=4pt]{$Q_6$}
            (Q2.E) 
                to [short] ++(0,-0.5) 
                to [R=$R_{E}$] ++(0,-0.7)
                node[ground]{}
            (Q5.B) 
                to[short,*-] ++(0,-0.7)
                to[short,-o] ++(-2.7,0)
                to[short,-*] ++(0,0) coordinate(LONinput)
                node[above]{$\mathrm{LO_N}$}
            (Q6.B)
                to[short] ++(0,-0.5)
                -| (Q3.B)
                to[short,*-*] ++(-0.4,0) coordinate(LOPinput)
                node[above]{$\mathrm{LO_P}$}
            (Q3.C)
                to[short,-*] ++(0,0.6) coordinate(Q3out)
                to[R=$R_{5}$] ++(0,1.2)
                node[vcc]{$V_{\mathrm{cc}}$}
            (Q4.C)
                to[short,-*] (Q6.C) coordinate(Q6out)
            (Q5.C)
                to[short,-] ++(0,0.6) coordinate(Q5out)
                to[short] (Q3out)
            (Q5out)
                to[short,*-o] ++(1.5,0) coordinate(OUTp)
                node[above]{$\mathrm{OUT_P}$}
            (Q6.C) 
                to[short,-o] ++(0.5,0) coordinate(OUTm)
                node[above]{$\mathrm{OUT_N}$}
            (Q4.C) 
                to[short,*-] ++(0,0.6)
                to[R=$R_{6}$] ++(0,1.2)
                node[vcc]{$V_{\mathrm{cc}}$}
            ;
            \draw 
            (IFNinput) 
                to[R=$R_{2}$] ++(-1.3,0)
                to[short,-*] ++(0,0.35) coordinate(biasIF)
                to[short] ++(0,0.35)
                to[R=$R_{1}$] (IFPinput)
            (LONinput) 
                to[R=$R_{4}$] ++(-1.28,0)
                to[short,-*] ++(0,0.35) coordinate(biasLO)
                to[short] ++(0,0.35)
                to[R=$R_{3}$] (LOPinput)
            (biasIF)
                to[R,-*] ++(-1,0) coordinate(biasIF_diodeconnect)
                to[R]    ++(-1,0)
                node[npn,anchor=B,xscale=-1] (Q7){}
                (Q7.center) node[anchor=east,xshift=4pt]{$Q_7$}
            (Q7.E)
                node[ground]{}
            (biasLO)
                to[R,-*] ++(-1,0) coordinate(biasLO_diodeconnect)
                to[R]    ++(-1,0)
                node[npn,anchor=B,xscale=-1] (Q8){}
                (Q8.center) node[anchor=east,xshift=4pt]{$Q_8$}
            (Q8.E)
                to[R] (Q7.C)
                to[short,-*] ++(0,-0.05)
                -| (biasIF_diodeconnect)
            (Q8.C) 
                to[short,*-] ++(0.1,0)
                -| (biasLO_diodeconnect)
            (Q8.C)
                to[R] ++(0,1) 
                node[vcc]{$V_{\mathrm{cc}}$};
            \draw[fill=white] (-2.5,2.8) rectangle (-0.5,4.2);
            
        \end{pgfonlayer}
        \begin{pgfonlayer}{foreground}
            \node at (-1.5,3.5) {
            \begin{tabular}{c|r}
                $R_{1,2}$ & $\SI{50}{\ohm}$\\
                $R_{3,4}$ & $\SI{55}{\ohm}$\\
                $R_{5,6}$ & $\SI{50}{\ohm}$\\
                $R_{E}$   & $\SI{5}{\ohm}$\\
            \end{tabular}
            };
        \end{pgfonlayer}
        \begin{pgfonlayer}{background}
            \fill[blue!7] (-4.5,-1.5) rectangle (-1.3,4);
            \node[anchor=west,black,align=right] at (-3.5,-1.1) {Current Mirror \\ Bias Circuit};
        \end{pgfonlayer}
    \end{circuitikz}
    \caption{Mixer schematic including bias circuitry. The resistors  $R_{1-6}$ function as termination resistors, creating differential virtual-ground nodes at the connections of $R_{1}$ and $R_2$, $R_{3}$ and $R_4$, and at $V_\mathrm{cc}$.}
    \label{fig:mixer_schem}
    \vspace{-0.2cm}
\end{figure}

\begin{figure}[t]
    \centering
    \begin{tikzpicture}
        \node[anchor=south] at (0,0) {\includegraphics[width=0.5\linewidth]{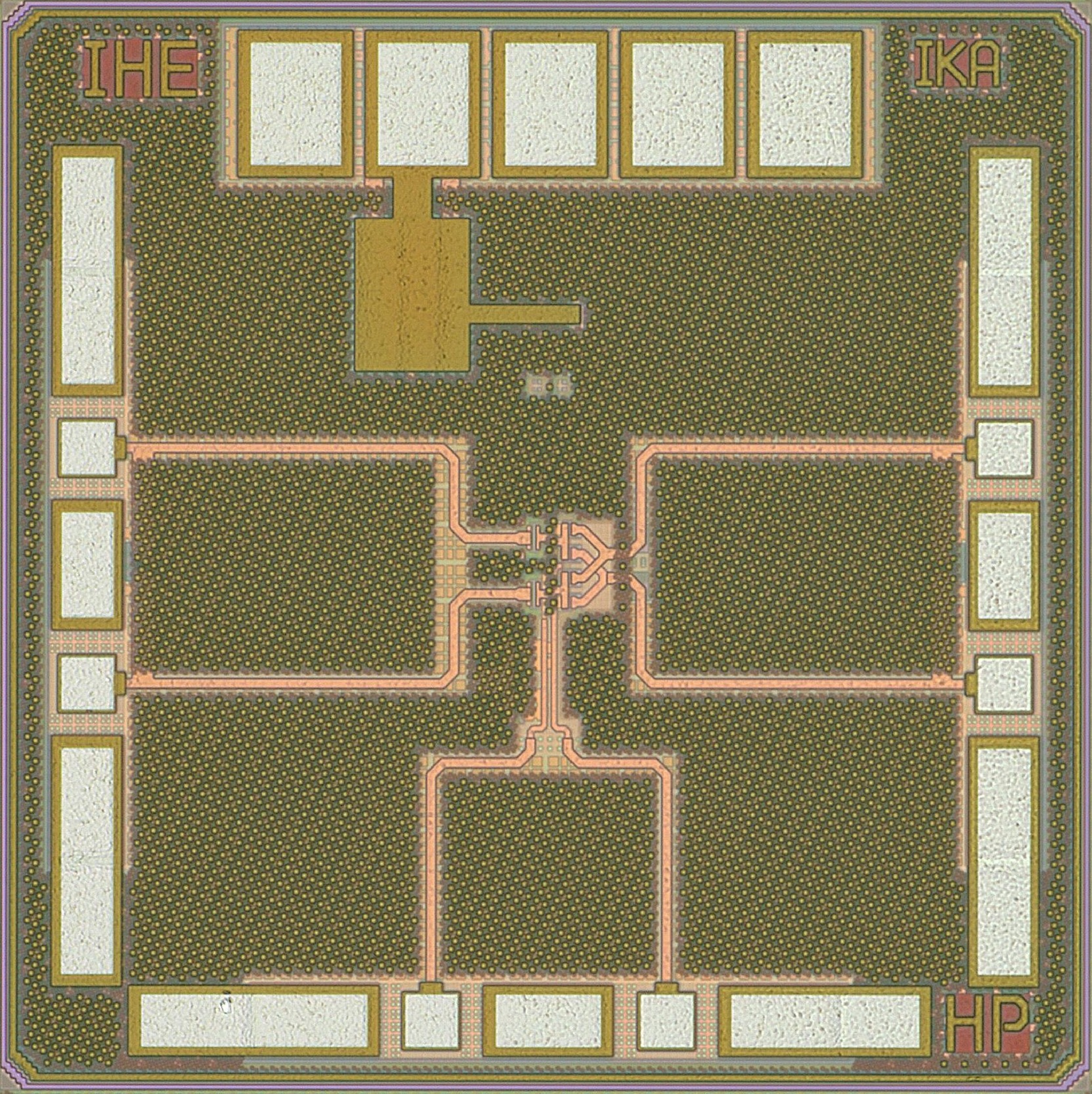}};
        \draw[red, very thick] (-0.3,1.7) rectangle (0.35,2.6);
        \node[draw=none,text width=0.5cm,text=white] at (-1.3,2.25) {IN};
        \node[draw=none,text width=0.5cm,text=white] at (1.3,2.25) {LO};
        \node[draw=none, text width=0.7cm, text=white] at (0,1) {OUT};
    \end{tikzpicture}
    \caption{Mixer micrograph, total size (including pads): $\qty{900}{\micro\meter}\times\qty{900}{\micro\meter}$, core area $\qty{100}{\micro\meter}\times\qty{180}{\micro\meter}$ highlighted in red.}
    \label{fig:mixer_micrograph}
    \vspace{-0.2cm}
\end{figure}

\section{System Design for Mixer Modeling}
\label{sec:system_model}

\begin{figure}[t]
    \centering
    \includegraphics[width=\linewidth]{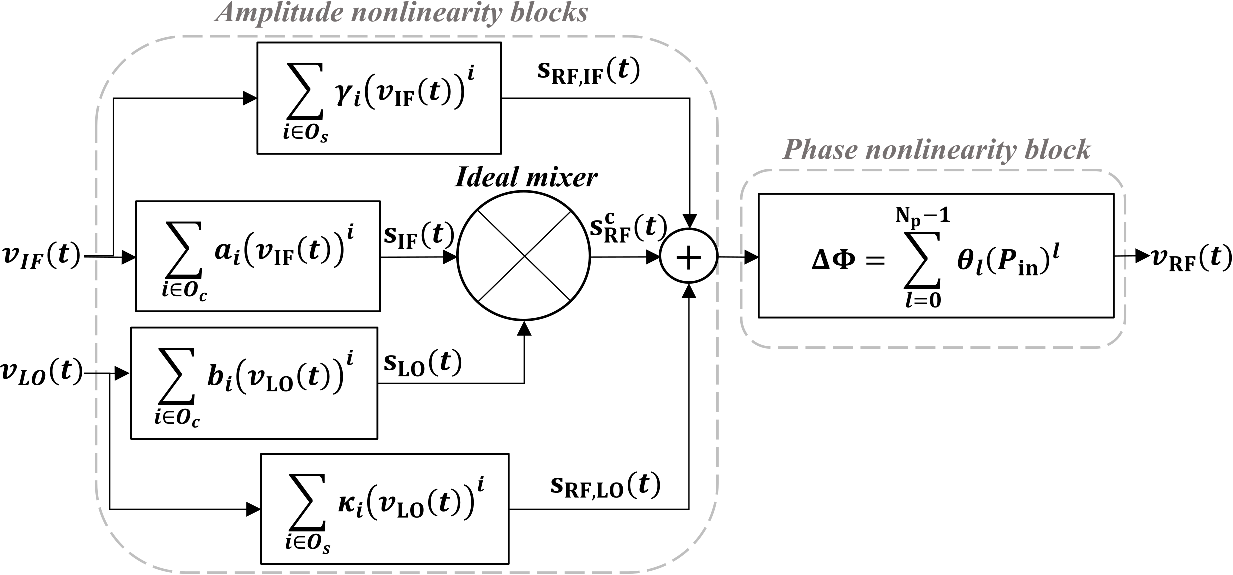}
    \caption{System model of the considered single-mixer multi-\ac{LO} architecture.}
    \label{fig:system_model}
    \vspace{-0.2cm}
\end{figure}
This work adopts a memoryless behavioral model for a double-balanced mixer when the mixer is simultaneously driven by multi-tone \ac{IF} and \ac{LO} signals. The input signal at port $\rho \in \{\text{IF}, \text{LO}\}$ is expressed as:
\begin{equation}
    v_\rho(t) = \sum_{i=1}^{N_\rho} A_{\rho,i} \cos(2\pi f_{\rho,i} t),
    \label{eq:multitone_input}
\end{equation}
where $N_\rho$ represents the number of tones, while $A_{\rho,i}$ and $f_{\rho,i}$ denote the amplitude and frequency of the $i$th tone at port $\rho$, respectively. 
Amplitude nonlinearities and \ac{IMD}s are captured by odd-order memoryless polynomials \cite{Multibox}
\begin{align}
s_{\mathrm{IF}}(t) &= \sum_{i\in\mathcal{O}_c} a_i \, (v_{\mathrm{IF}}(t))^i, \quad s_{\mathrm{LO}}(t) = \sum_{i\in\mathcal{O}_c} b_i \, (v_{\mathrm{LO}}(t))^i,
\label{eq:core1}
\end{align}
where \( \mathcal{O}_c = \{1, 3, \dots, 2\mathrm{K}_{\mathrm{c}}-1\} \), with $2\mathrm{K}_{\mathrm{c}}-1$ denoting the maximum nonlinearity order of core amplitude blocks and \( \boldsymbol{\alpha} = [\mathbf{a}; \mathbf{b}] \in \mathbb{R}^{2\mathrm{K}_\mathrm{c}} \) are the amplitude nonlinearity coefficients. For simplicity, the same order is assumed for core blocks.

The ideal mixing core shown in Fig.~\ref{fig:system_model} is a pure multiplier with nonlinear blocks. Therefore, the nonlinear RF output (before phase compensation) is modeled as:
\begin{equation}
s_\text{RF}(t) = s_\text{IF}(t) \cdot s_\text{LO}(t) + s_{\text{RF,IF}}(t) + s_{\text{RF,LO}}(t),
\end{equation}
where \(s_{\text{RF,IF}}(t)\) and \(s_{\text{RF,LO}}(t)\) model sidebranch distortion at the \ac{IF} and \ac{LO} ports using separate odd-order polynomials with coefficients \(\gamma_i\) and \(\kappa_i\), respectively. These sidebranches capture leakages (e.g., \(f_{\text{IF},i}\)), harmonics (e.g., \(3f_{\text{LO},i}\)), intra-port IMDs (e.g., \(2f_{\text{LO},1} \pm f_{\text{LO},2}\)), distinct from core \ac{IF}–\ac{LO} intermodulation. \( \mathcal{O}_s = \{1, 3, \dots, 2\mathrm{K}_{\text{s}}-1\} \) is the sidebranch amplitude nonlinearity orders with $2\mathrm{K}_{\mathrm{s}}-1$ denoting the maximum nonlinearity order of sidebranch amplitude blocks. These distortion terms are modeled using additional polynomial blocks and fitted in a dedicated sidebranch estimation step (see Algorithm \ref{alg:1}).


Phase nonlinearity is applied to dominant fundamental bins using an input power-dependent polynomial:
\begin{equation}
\Delta \Phi(P_{\text{in}})=\sum_{l=0}^{\mathrm{N_p}-1}\theta_l (P_\text{in})^{l}.
\label{eq:phase_nonlinearity}
\end{equation}
where $\mathrm{N_p}$ and $P_\text{in}$ denote the order of polynomial and input power at \ac{IF} port. In this study, phase nonlinearity is applied only to fundamental mix products, which is sufficient to represent near-saturation behavior.

The behavioral model is optimized by fitting its predicted amplitude responses to reference curves obtained from an input power sweep. Specifically, we fit the predicted fundamental \ac{AM-AM} response \( f_{\text{F}}(\boldsymbol{\alpha}) \) and \ac{IM3} response \( f_{\text{IM3}}(\boldsymbol{\alpha}) \) to their reference counterparts \( y_{\text{F}} \) and \( y_{\text{IM3}} \), while simultaneously minimizing spectral-domain error across selected frequency bins, where spectral loss is computed at the reference input power \( P_{\text{in}}^\text{ref} \).
One-sided complex spectrum of the measured output and the corresponding power spectrum in dBm are defined as $Y_{\text{r}} \in \mathbb{C}^M$ and $S_{\text{r}} \in \mathbb{R}^M$, where $M$ denotes the number of \ac{DFT} bins.

\textbf{Core Fitting:} First, a set of frequency bin indices $\mathcal{B}_{\text{c}} \subset \{1, \dots, M\}$ is found by identifying the theoretical frequency bins of odd-order mixing products in $S_{\text{r}}$. The indices are partitioned into strong ($B_\text{s}$) and weak ($B_\text{w}$) sets based on thresholds $\tau_\text{w}$ and $\tau_\text{s}$:
\begin{equation}
    B_s = \{b \in \mathcal{B}_{\text{c}} \mid S_{\text{r}}[b] \ge \tau_s \}, \quad B_w = \{b \in \mathcal{B}_{\text{c}} \mid \tau_w \le S_{\text{r}}[b] < \tau_s \}.
    \label{eq:bins}
\end{equation}
In \eqref{eq:bins}, $\tau_\text{w}$ is used to reduce the number of bins and relax the optimization problem.
Optionally, frequency-based limitation can be used to reduce number of bins. Using these sets, the spectral loss is defined as:
\begin{equation}
    \mathcal{L}_{\text{s}}(\boldsymbol{\alpha}) = w_s \left\| S_{\text{p}} - S_{\text{r}} \right\|_{2, B_s} + w_w \left\| \max(\mathbf{0}, S_{\text{p}} - S_{\text{r}}) \right\|_{2, B_w},
    \label{eq:spec_loss}
\end{equation}
where $\ \|\cdot\|_{2, B}$ ,$S_\text{p}$, $w_\text{s}$ and $w_\text{w}$ denote the Euclidean norm over set $B$, predicted one-sided spectrum in dBm, the weights for strong and weak bins, respectively. 
The parameters \( \boldsymbol{\alpha} = [\mathbf{a}; \mathbf{b}] \in \mathbb{R}^{2\mathrm{K}_\mathrm{c}} \) are identified by solving:
\begin{align}
\mathrm{Obj}(\boldsymbol{\alpha}) 
&= w_{\text{F}} \cdot \| f_{\text{F}}(\boldsymbol{\alpha}) - y_{\text{F}} \|_2 \nonumber \\  &+w_{\text{IM3}} \cdot \| f_{\text{IM3}}(\boldsymbol{\alpha}) - y_{\text{IM3}} \|_2  
+ \mathcal{L}_{\text{s}}(\boldsymbol{\alpha}), 
\label{eq:core_opt_obj1} \\[0.4em]
&\quad \boldsymbol{\alpha}^\mathrm{opt} = \arg\min_{\boldsymbol{\alpha} \in \Theta} \; 
\mathrm{Obj}(\boldsymbol{\alpha}).
\label{eq:core_opt_obj2}
\end{align}
In \eqref{eq:core_opt_obj1}, $w_\text{F}$ and $w_\text{IM3}$ are the weights for fundamental and IM3 curve fitting. The feasible set \( \Theta \) imposes box constraints: \( l_i \leq \alpha_i \leq u_i \). Optionally, regularization can be applied to keep high-order polynomial coefficients bounded. The optimization is solved via multi-start based search.

\textbf{Sidebranch Fitting:} For the sidebranches, a second set of indices $\mathcal{B}_{\text{sb}} \subset \{1, \dots, M\}$ is identified based on theoretical intra \ac{IMD}, harmonic and leakage frequencies. The error of the complex sum of core and sidebranches is minimized by:
\begin{equation}
    \min_{\boldsymbol{\gamma}, \boldsymbol{\kappa}} \left\| \mathcal{P}_{\text{dBm}}\big( Y_{\text{c}} + Y_{\text{sb}}(\boldsymbol{\gamma}, \boldsymbol{\kappa}) \big) - S_{\text{r}} \right\|_{2, \mathcal{B}_{\text{sb}}},
    \label{eq:sb_fitting}
\end{equation}
where $Y_{\text{c}}$ is the fixed complex spectrum from the optimized core, and $Y_{\text{sb}}$ is the complex spectrum of the sidebranches, $\mathcal{P}_{\text{dBm}}(\cdot)$ converts the complex sum to dBm.

\textbf{Phase Fitting:} Finally, phase nonlinearity is applied based on \eqref{eq:phase_nonlinearity}. Specifically for each fundamental bin $ k \in \{1,\dots,N_\text{fund}\} $, phase nonlinearity curves are extracted from corresponding bins across input powers and fitted via \eqref{eq:phase_nonlinearity}.

All optimization steps are summarized in Algorithm~\ref{alg:1}. 

\begin{algorithm}[h!]
\caption{Multi-tone Mixer Model Fitting}
\begin{algorithmic}[1]
\Require \( y_\text{F}, y_\text{IM3}, S_\text{r}, \tau_\mathrm{s}, \tau_\mathrm{w},\Theta \), \( \mathcal{B}_{\text{c}} \), \( \mathcal{B}_{\text{sb}} \), $w_{\text{F}}$, $w_{\text{IM3}}$, $w_{\text{s}}$, $w_{\text{w}}$
\Ensure Optimal coefficients: \(\boldsymbol{\alpha}^{\mathrm{opt}}, \boldsymbol{\gamma}^{\mathrm{opt}}, \boldsymbol{\kappa}^{\mathrm{opt}} \)

\State \textbf{Step 1: Amplitude Core Fitting}
\For{$s = 1$ to $N_{\text{starts}}$}
  \State Initialize \( \boldsymbol{\alpha}_s = [\mathbf{a}, \mathbf{b}] \in \Theta \)
  \State Minimize the objective in~\eqref{eq:core_opt_obj2}
\EndFor
\State Select \( \boldsymbol{\alpha}^{\mathrm{opt}} = \arg\min \text{Obj}(\boldsymbol{\alpha}_s) \)

\State \textbf{Step 2: Sidebranch Fitting} 
\State Minimize the objective in~\eqref{eq:sb_fitting}
\State \textbf{Step 3: Phase Fitting}
\For{$k = 1$ to $N_{\text{fund}}$}
  \State Fit \( \Delta \phi_k(P_{\text{in}}) \approx \sum_l \theta_{k,l} P_{\text{in}}^l \)
  \State Apply phase rotation on bin $k$ based on \(\Delta \phi_k(P_\text{in}) \)
\EndFor
\end{algorithmic}
\label{alg:1}
\end{algorithm}


\section{Measurement Results}
\label{meas_results}

\begin{figure}[b]
\centering
    \psfrag{XXX1}[c][c]{\scriptsize $\mathrm{P_{in}~(dBm)}$}
	\psfrag{YYY1}[c][c]{\scriptsize $\mathrm{P_{out}^{fund}~(dBm)}$}
    \psfrag{XXX2}[c][c]{\scriptsize $\mathrm{P_{in}~(dBm)}$}
	\psfrag{YYY2}[c][c]{\scriptsize $\mathrm{P_{out}^{IM3}~(dBm)}$}
    \psfrag{(a)}[c][c]{\footnotesize (a)}
    \psfrag{(b)}[c][c]{\footnotesize (b)}
	\includegraphics[width=0.85\linewidth]{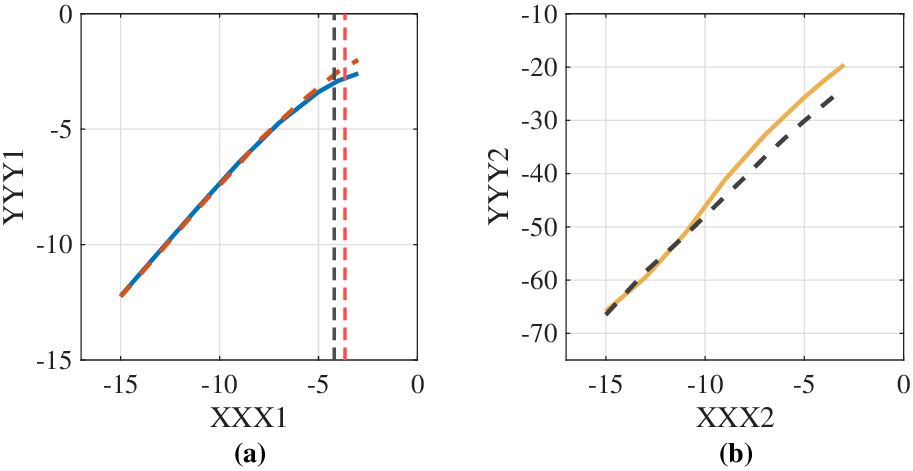}
	\captionsetup{justification=raggedright,labelsep=period,singlelinecheck=false}
	\caption{\ ADS ({\color[rgb]{0,0.451,0.741}\textbf{\rule[0.5ex]{1.25em}{1.25pt}}}) and measurement ({\color[rgb]{0.8471,0.3216,0.0941}\textbf{\rule[0.5ex]{0.75em}{1.25pt}~\rule[0.5ex]{0.25em}{1.25pt}}}) comparison for fundamental mix ($\mathrm{f_{IF,1}+f_{LO}}$) in (a) for IM3 ($\mathrm{2f_{IF,1}-f_{IF,2}+f_{LO}}$) ({\color[rgb]{0.9294,0.6941,0.3137}\textbf{\rule[0.5ex]{1.25em}{1.25pt}}}), ({\color[rgb]{0.2510,0.2510,0.2510}\textbf{\rule[0.5ex]{0.75em}{1.25pt}~\rule[0.5ex]{0.25em}{1.25pt}}}) in (b).
    $\mathrm{P}_{\qty{1}{dB}}$ for each ({\color[rgb]{0,0,0}\textbf{\rule[0.5ex]{0.75em}{1.25pt}~\rule[0.5ex]{0.25em}{1.25pt}}}), ({\color[rgb]{1,0,0}\textbf{\rule[0.5ex]{0.75em}{1.25pt}~\rule[0.5ex]{0.25em}{1.25pt}}}) is indicated in (a) respectively.}

	\vspace{-0.2cm}
    \label{fig:ADS_vs_meas}
\end{figure}

Before validating the mixer model with multi-tone excitation, simulations are verified against measurements of the Gilbert-cell mixer. 
Parameters are listed in Table~\ref{tab:char_par}.
\begin{table}[t]
\vspace{-0.2cm}
    \renewcommand{\arraystretch}{1.5}
    \setlength{\arrayrulewidth}{.1mm}
    \setlength{\tabcolsep}{4pt}
    		
    \centering
    \captionsetup{width=43pc,justification=centering,labelsep=newline}
    \caption{\textsc{Considered System Parameters for Characterization}}
    \label{tab:char_par}
    \begin{tabular}{|c|c|c|c|}
        \hhline{|====|}
         & $\textbf{$f_{\mathrm{IF}} $}$  (GHz) & $\textbf{$f_{\mathrm{LO}}$}$ (GHz) & $\textbf{$P_{\mathrm{LO}}$}$ (dBm) \\\hline
        \textbf{Single-tone-LO (meas.)} &$0.995,1.005$ & $9$ & 0  \\\hline
        \textbf{Multi-tone-LO (sim.)} & $0.995,1.005$ & $9,9.5$ & 0 per tone
        \\\hhline{|====|}
    \end{tabular}
    \vspace{-0.2cm}
\end{table}
The simulated and measured \ac{AM-AM} characteristics for the single-tone \ac{LO} case are compared in Fig.~\ref{fig:ADS_vs_meas}, which highlights a good agreement and demonstrates the ability to predict nonlinear circuit behavior accurately in simulations.


\begin{figure*}[t]
    \centering
    \psfrag{5}[c][c]{\tiny $5$}
    \psfrag{6}[c][c]{\tiny $6$}
    \psfrag{7}[c][c]{\tiny $7$}
    \psfrag{8}[c][c]{\tiny $8$}
    \psfrag{9}[c][c]{\tiny $9$}
    \psfrag{10}[c][c]{\tiny $10$}
    \psfrag{11}[c][c]{\tiny $11$}
    \psfrag{12}[c][c]{\tiny $12$}
    \psfrag{13}[c][c]{\tiny $13$}
    \psfrag{-4}[c][c]{\tiny -$4$}
    \psfrag{-6}[c][c]{\tiny -$6$}
    \psfrag{-8}[c][c]{\tiny -$8$}
    \psfrag{-10}[c][c]{\tiny -$10$}
    \psfrag{-12}[c][c]{\tiny -$12$}
    \psfrag{-14}[c][c]{\tiny -$14$}
    \psfrag{-16}[c][c]{\tiny -$16$}
    \psfrag{-2}[c][c]{\tiny -$2$}
    \psfrag{0}[c][c]{\tiny $0$}
    \psfrag{-55}[c][c]{\tiny -$55$}
    \psfrag{-50}[c][c]{\tiny -$50$}
    \psfrag{-45}[c][c]{\tiny -$45$}
    \psfrag{-40}[c][c]{\tiny -$40$}
    \psfrag{-35}[c][c]{\tiny -$35$}	
    \psfrag{-25}[c][c]{\tiny -$25$}	
     \psfrag{-60}[c][c]{\tiny -$60$}
    \psfrag{-20}[c][c]{\tiny -$20$}
    \psfrag{-30}[c][c]{\tiny -$30$}	       
    \psfrag{XXX1}[c][c]{\tiny $\mathrm{P_{in}~(dBm)}$}
	\psfrag{YYY1}[c][c]{\tiny $\mathrm{P_{out}^{fund}~(dBm)}$}
    \psfrag{XXX2}[c][c]{\tiny $\mathrm{P_{in}~(dBm)}$}
	\psfrag{YYY2}[c][c]{\tiny $\mathrm{P_{out}^{IM3}~(dBm)}$}
    \psfrag{XXX3}[c][c]{\tiny $\mathrm{Frequency~(GHz)}$}
	\psfrag{YYY3}[c][c]{\tiny $\mathrm{Power~(dBm)}$}
    \psfrag{(a)}[c][c]{\scriptsize (a)}
    \psfrag{(b)}[c][c]{\scriptsize (b)}
    \psfrag{(c)}[c][c]{\scriptsize (c)}
    \includegraphics[width=0.8\textwidth]{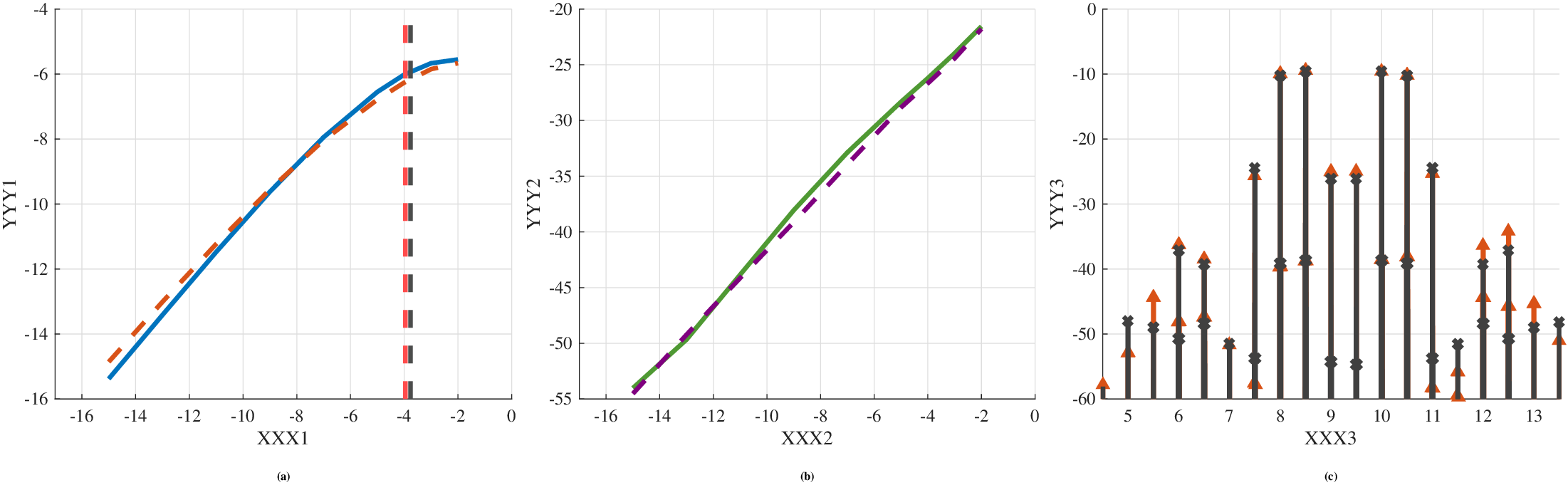}
    \caption{\ \ac{AM-AM} curve of fundamental mix (${f_\mathrm{IF1}+f_\mathrm{LO1}}$) for \ac{ADS} ({\color[rgb]{0,0.451,0.741}\textbf{\rule[0.5ex]{1.25em}{1.25pt}}}) and multibox model ({\color[rgb]{0.8471,0.3216,0.0941}\textbf{\rule[0.5ex]{0.75em}{1.25pt}~\rule[0.5ex]{0.25em}{1.25pt}}}) with ($\mathrm{P}_{\qty{1}{dB}}$)  for ADS ({\color[rgb]{1,0,0}\textbf{\rule[0.5ex]{0.75em}{1.25pt}~\rule[0.5ex]{0.25em}{1.25pt}}}) and multibox ({\color[rgb]{0,0,0}\textbf{\rule[0.5ex]{0.75em}{1.25pt}~\rule[0.5ex]{0.25em}{1.25pt}}}) indicated. \ac{IM3} (${2f_\mathrm{IF1}-f_\mathrm{IF2}+f_\mathrm{LO1}}$) \ac{AM-AM} results for \ac{ADS} ({\color[rgb]{0.3059,0.6000,0.2039}\textbf{\rule[0.5ex]{1.25em}{1.25pt}}}) and multibox ({\color[rgb]{0.5020,0,0.5020}\textbf{\rule[0.5ex]{0.75em}{1.25pt}~\rule[0.5ex]{0.25em}{1.25pt}}}) are shown in (b). The spectrums for \ac{ADS} ({\color[rgb]{0.8471,0.3216,0.0941}\textbf{\rule[0.5ex]{1.25em}{1.25pt}}}) and multibox model ({\color[rgb]{0.2510,0.2510,0.2510}\textbf{\rule[0.5ex]{1.25em}{1.25pt}}}) for 4--\qty{14}{\giga\hertz} are shown in (c).}
   \label{fig:full_width_top}
   \vspace{-0.2cm}
\end{figure*}

Multi-tone \ac{LO} characterization could not be performed due to experimental setup limitations, so ADS results are used as a reference for the multibox model comparison. Multi-tone \ac{LO} parameters are listed in Table~\ref{tab:char_par} and ADS--multibox model comparisons are shown in Fig.~\ref{fig:full_width_top}. A reduction in conversion gain is observed in Fig.~\ref{fig:full_width_top}a for multi-tone \ac{LO} excitation compared to the single-tone case, despite identical per-tone power (Table~\ref{tab:char_par}). This is attributed to reduced mixer switching efficiency caused by the addition of multiple \ac{LO} tones, as constructive and destructive addition of tones changes the \ac{LO} waveform's drive symmetry.

\begin{figure}[t]
\centering
    \psfrag{XXX1}[c][c]{\scriptsize $\mathrm{Doppler~(kHz)}$}
	\psfrag{YYY1}[c][c]{\scriptsize $\mathrm{Range~(m)}$}
    \psfrag{ZZZ1}[c][c]{\scriptsize $\mathrm{Normalized~Power~(dB)}$}
    \psfrag{XXX2}[c][c]{\scriptsize $\mathrm{Doppler~(kHz)}$}
	\psfrag{YYY2}[c][c]{\scriptsize $\mathrm{Range~(m)}$}
    \psfrag{ZZZ2}[c][c]{\scriptsize $\mathrm{Normalized~Power~(dB)}$}
    \psfrag{(a)}[c][c]{\footnotesize (a)}
    \psfrag{(b)}[c][c]{\footnotesize (b)}
    \includegraphics[width=0.85\linewidth]{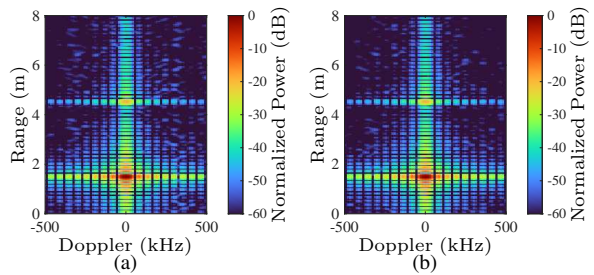}
    \caption{Simulated range-Doppler maps: (a) ADS and (b) multi-box model based on the parameters in Table~\ref{tab:OFDM_par}.}
    \label{fig:RDM}
    \vspace{-0.2cm}
\end{figure}

The sensing performance of ADS and the proposed model is compared by upconverting a frequency comb OFDM waveform (centered at \qty{1}{\giga\hertz}) to \qty{10}{\giga\hertz} and \qty{10.5}{\giga\hertz}, and downconverting the resulting signals to baseband. The parameters are given in Table~\ref{tab:OFDM_par}.
\begin{table}[t]
\vspace{-0.2cm}
    \renewcommand{\arraystretch}{1.5}
    \setlength{\arrayrulewidth}{.1mm}
    \setlength{\tabcolsep}{4pt}   		
    \centering   \captionsetup{width=43pc,justification=centering,labelsep=newline}
    \caption{\textsc{Considered System Parameters}}
    \label{tab:OFDM_par}
    \begin{tabular}{|c|c|}
        \hhline{|==|}
        \textbf{Carrier frequency at \ac{IF}  ($f_\mathrm{c}$)} & \qty{1}{\giga\hertz} \\\hline
        \textbf{\ac{LO} tone frequencies ($f_\mathrm{LO}$)} & \SI{9}{\giga\hertz}, \qty{9.5}{\giga\hertz} \\\hline
        \textbf{Frequency bandwidth ($B$)} & \qty{500}{\mega\hertz} \\\hline
        \textbf{Number of subcarriers ($N$)} & \num{256} \\\hline
        \textbf{Cyclic prefix length ($N_{\mathrm{cp}}$)} & $32$ \\\hline
        \textbf{OFDM symbols per frame ($M$)} & \num{32} \\\hline
        \textbf{Average \ac{IF}  signal power ($P_{\mathrm{Tx}}$)} & \qty{-13}{dBm} \\\hline
        \textbf{PAPR} & \qty{12.53}{dB} \\\hhline{|==|}
    \end{tabular}
    \vspace{-0.2cm}
\end{table}  
Fig.~\ref{fig:RDM} shows radar images for a two-target scenario. The overlap between the 3rd-order \ac{LO} intra-\ac{IMD} and the upconverted OFDM signal at 10 GHz causes distortion in the central subcarriers.
To mitigate this, we suggest cancelling subcarriers around the center for communication purposes. Optionally, increasing the \ac{IF}  frequency or reducing the \ac{LO} tone power were not used in this study to preserve the frequency plan and avoid degrading the \ac{LO} drive. The impact on sensing is assumed tolerable.
 
For insufficient input backoff, both models lead to noisy radar images. To evaluate radar image quality, we define the image \ac{SINR} as the ratio of the power of the stronger target to the average noise floor in the radar image, calculated after masking the strong range and Doppler sidelobes of both targets. Additionally, \ac{PSLR} and \ac{ISLR} are computed along the Doppler axis for the stronger target.
The multibox model achieves ${\mathrm{SINR} = \qty{61.86}{dB}}$, ${\mathrm{PSLR} = \qty{-13.41}{dB}}$, and ${\mathrm{ISLR} = \qty{-9.71}{dB}}$--closely matching with \ac{ADS} results which are ${\mathrm{SINR} = \qty{60.96}{dB}}$, ${\mathrm{PSLR} = \qty{-13.42}{dB}}$, and ${\mathrm{ISLR} = \qty{-9.70}{dB}}$.
The difference in $\mathrm{SINR}$ can be attributed to phase nonlinearity modeling. Phase nonlinearity is applied uniformly to all subcarriers, simplifying model fitting. While this avoids scenario-dependent complexity of per-subcarrier phase nonlinearity, it overestimates and slightly boosts $\mathrm{SINR}$. To eliminate this deviation, subcarrier-wise phase correction can be incorporated for improved accuracy.

\section{Conclusion}
In this paper, an optimization-based behavioral modeling framework is introduced for mixers operating under multi-tone \ac{LO} excitation, specifically to enable efficient and fast system-level performance assessment and waveform design for frequency comb OFDM radar applications. By jointly modeling amplitude nonlinearities, \ac{IMD}, harmonics, and phase distortion in the spectral domain, the framework accurately captures key non-idealities. Spectrum-aware fitting enables precise modeling of strong and weak bins, while a simplified phase nonlinearity block for all subcarriers avoids complex subcarrier-dependent correction. Validation against ADS simulations shows a strong agreement in nonlinear behavior and sensing performance in image \ac{SINR}, PSLR, and ISLR. 

\section*{Acknowledgment}
The authors acknowledge the financial support by the German Research Foundation in “Broadband OFDM Radar System at Reduced Sampling Rates Using Frequency Combs (BORA)” (grant number: 467104011).

\bibliographystyle{IEEEtran}
\bibliography{references}

@INPROCEEDINGS{survey_ISAC,
  author={{Mandelli et al.},S.},
  booktitle={Proc. 6GNet}, 
  title={Survey on Integrated Sensing and Communication Performance Modeling and Use Cases Feasibility}, 
  year={2023},
  volume={},
  number={},
  pages={1-8},
  doi={10.1109/6GNet58894.2023.10317691}}

@article{Multibox,
author = {{Ozgun et al.},A.},
title = {A Multi-Box Behavioural Nonlinear Mixer Model},
journal = {International Journal of Electronics},
volume = {111},
number = {8},
pages = {1384--1402},
year = {2024},
publisher = {Taylor \& Francis},
doi = {10.1080/00207217.2023.2233120}
}

@ARTICLE{Nuss_comb,
  author={{Nuss et al.}, B.},
  journal={IEEE Trans. Microw. Theory Tech.}, 
  title={Frequency Comb {OFDM} Radar System With High Range Resolution and Low Sampling Rate}, 
  year={2020},
  volume={68},
  number={9},
  pages={3861-3871},
  doi={10.1109/TMTT.2020.2988254}}

@INPROCEEDINGS{Vandermot,
  author={{Vandermot et al.},K.},
  booktitle={IEEE Instrum. Meas. Technol. Conf. (IMTC)}, 
  title={Understanding the Nonlinearity of a Mixer Using Multisine Excitations}, 
  year={2006},
  volume={},
  number={},
  pages={1205-1209},
  doi={10.1109/IMTC.2006.328451}}

@INPROCEEDINGS{Aksoyak2024,
author = {{Aksoyak et al.},{\.I}. K.}, 
  booktitle={IEEE SiRF}, 
  title={A {SiGe}-Based Quadrature {D-Band} Up-Converter with High Output Power}, 
  year={2024},
  volume={},
  number={},
  pages={33-36},
  doi={10.1109/SiRF59913.2024.10438550},
  doi={10.1109/SiRF59913.2024.10438550}
}

\end{document}